\def\year{2019}\relax
\documentclass[letterpaper]{article} 
\usepackage{aaai19}  
\usepackage{times}  
\usepackage{helvet}  
\usepackage{courier}  
\usepackage{url}  
\usepackage{graphicx} 
\usepackage{xcolor}

\usepackage{amsmath}
\begin{document}
%
\title{Founding The Domain of AI Forensics}
\author{Ibrahim Baggili and 
 Vahid Behzadan\\ 
 University of New Haven\\
 ibaggili@newhaven.edu,
 vbehzadan@newhaven.edu}

\maketitle
\begin{abstract}
 With the widespread integration of AI in everyday and critical technologies, it seems inevitable to witness increasing instances of failure in AI systems. In such cases, there arises a need for technical investigations that produce legally acceptable and scientifically indisputable findings and conclusions on the causes of such failures. Inspired by the domain of cyber forensics, this paper introduces the need for the establishment of \emph{AI Forensics} as a new discipline under AI safety. Furthermore, we propose a taxonomy of the subfields under this discipline, and present a discussion on the foundational challenges that lay ahead of this new research area. 
\end{abstract}

\section{Introduction}
Recent advances in Artificial Intelligence (AI) have given rise to the rapidly growing adoption of such techniques by a vast array of industries and technologies. The penetration of AI in our day-to-day lives is easily observed in everyday technologies such as advertisement and road navigation (e.g., Google Maps), as well as critical sectors such as cybersecurity \cite{li2018cyber}, healthcare \cite{jiang2017artificial}, and smart cities \cite{mckee2018survey}. However, the growing complexity of AI techniques renders the assurance and verification of safety and reliability of such systems difficult \cite{yampolskiy2018artificial}. Therefore, it is not surprising to observe the growing frequency of reported failures in AI-enabled systems (e.g., \cite{yampolskiy2016artificial}). 

In response, the evolving field of \emph{AI safety} \cite{amodei2016concrete} aims to tackle the problem of reliability and safety in AI-enabled systems. The resulting body of work to date is largely focused on the prevention of unsafe behavior in current and future AI technologies. However, the rapid penetration of AI into critical technologies has greatly outpaced the research efforts of the AI safety community. Hence, an increase in the frequency of failures in deployed AI seems inevitable. In the event of such failures in critical systems, it becomes necessary to investigate the causes and sequence of events leading to the failure. Besides the analysis of underlying technical deficiencies, such investigations will need to determine a variety of other aspects, including: whether the failure has been the result of malicious actions, which party is liable for the damages caused by the failure, and whether the failure could have been prevented. Furthermore, interested parties such as law enforcement and insurance providers may require this investigation to result in legally acceptable and indisputable findings and conclusions.   

Similar needs in the domain of computer safety and security have given rise to the field of \emph{Cyber Forensics}. Digital Forensics, also known as Cyber Forensics, revolves around the scientific and legal extraction of digital evidence. This field is multidisciplinary and involves computing, law, criminology, psychology and other disciplines. At the core of the domain, however, is the Acquisition, Authentication and Analysis (AAA) of digital evidence. 


Inspired by this analogue, we argue for the need to establish the formalism of \emph{AI Forensics} as a new discipline under AI safety. This formalism will aim to develop the tools, techniques, and protocols for the forensic analysis of AI failures. Accordingly, this paper makes the following contributions:

\begin{itemize}
    \item We offer the first working definition of AI Forensics.
    \item We conceptualize the first formal attempt of the AI Forensics domain, and propose a taxonomy for the corresponding types and sources of evidence.
    \item We enumerate a number of notable challenges in the domain of AI Forensics.  
\end{itemize}







\begin{figure*}[t]
	\includegraphics[width = 0.9\textwidth]{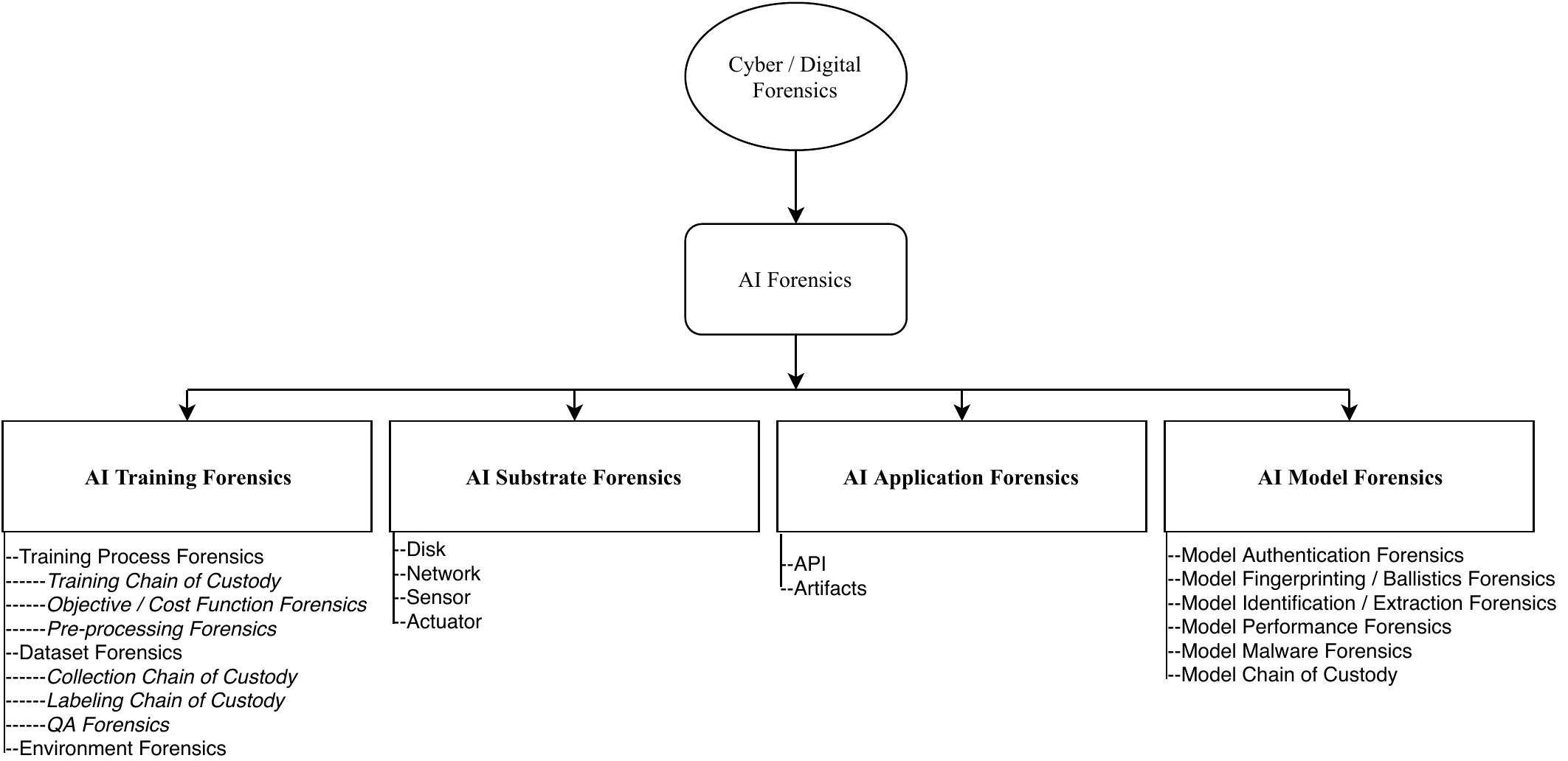}
	\caption{Diagram of the AI Forensics Research Domain and Sub-Domains}
	\label{fig:aiforensics}
\end{figure*}

\section{Related Work}
In recent years, there has been a growing interest in failure detection and analysis techniques for algorithmic decision making\cite{goodman2016eu}. This is partly due to the European Union's General Data Protection Regulations (GDPR), which requires the explainability of consequential decisions made by algorithms. Similarly, the research on Explainable AI \cite{samek2017explainable} aims to create tools and techniques that enable the explainability of black-box models such as deep neural networks. However, current state of the art in the analysis of failures in AI, and in particular machine learning models, is largely focused on the technical diagnosis and troubleshooting of design and training issues (e.g., \cite{nushi2018towards}). Hence, there remains a gap with regards to tools and techniques that enable the forensic analysis of failures in AI-enabled systems. 


\section{Digital Forensics and Digital Evidence}
 Digital forensics is defined as \emph{ ``The use of scientifically derived and proven methods toward the preservation, collection, validation, identification, analysis, interpretation, documentation and presentation of digital evidence derived from digital sources for the purpose of facilitating or furthering the reconstruction of events found to be criminal, or helping to anticipate unauthorized actions shown to be disruptive to planned operations''} \cite{palmer2001road}. An important component of digital evidence is its admissibility to the court of law. 

The admissibility of evidence was mostly dominated by what is known as the Frye test that articulated expert scientific evidence is admissible only if the scientific community generally accepts it. Since 1993, courts in the United States have adopted Rule 702 of the federal rules of evidence, resulting in what many refer to as the \emph{Daubert process}. This process comprises of four major guidelines \cite{daubert}:

\begin{enumerate}
    \item Testing: Can and has the procedure been tested?
    \item Error Rate: Is there a known error rate of the procedure?
    \item Publication: Has the procedure been published and subject to peer review?
    \item Acceptance: Is the procedure generally accepted in the relevant scientific community?
\end{enumerate}

Digital forensics has made strides by exploring areas such as disk forensics \cite{disk1}, memory forensics \cite{memory1}, network forensics \cite{network1}, cloud forensics \cite{cloud1}, artifact forensics \cite{artifact2}, blockchain storage and cryptocurrency forensics \cite{blockchain1}, social media forensics \cite{social3}, authorship attribution \cite{authorship1}, and mobile forensics \cite{mobile3}.

The aforementioned forensic areas also benefit from AI techniques, as process automation in digital forensics is of importance given the volume, variety and velocity production of data.
While the opportunity for AI Forensics has been recently noted by experts \cite{luciano2018digital}, at the time of writing this work, the domain was ill-defined. 
\section{The Landscape of AI Forensics}
We view the scope of AI Forensics as a subfield of digital forensics, defined as the \emph{scientific and legal tools, techniques, and protocols for the extraction, collection, analysis, and reporting of digital evidence pertaining to failures in AI-enabled systems.} In compliance with the Daubert process, AI Forensics provides a framework to enable the systematic and scientific resolution of such questions as:
\begin{itemize}
    \item What were the sequence of events and conditions that led to the failure?
    \item Did the failure result from malicious actions?
    \item Which party or parties is responsible for the failure?
    \item Would it have been possible to prevent the failure?
    \item Where did the failures take place?
\end{itemize}

The core of any forensic investigation is the collection and extraction of evidence. To this end, we introduce a classification of the various types evidence that can be of relevance to the investigation. Furthermore, we identify possible sources of evidence for each of the enumerated types.

\subsection{AI Training Forensics}
In forensic investigations of machine learning systems, a necessary step is to identify potential intentional or unintentional faults introduced during the training of the system. Such faults may stem from any of the components involved in training, as detailed below:  
\subsubsection{Training Process Forensics:}
The training process comprises of the optimization algorithm and its corresponding hyperparameters. A major cause of AI failures is the misspecification of the objective or cost function, which may result in behaviors that are misaligned with the goals of the designer \cite{arnold2017value}. Furthermore, design choices such as exploration techniques in reinforcement learning agents \cite{behzadan2018faults} and regularization techniques may result in brittle behaviors that fail to adapt to distributional shifts in their settings of deployment \cite{papernot2018sok}. Therefore, such parameters and choices constitute valuable forensic evidence.

\subsubsection{Dataset Forensics:}
The dataset used in the training of a machine learning model may be of inconsistent or unrepresentative samples, thus resulting in a model that is not compatible with the conditions of its deployment settings. Furthermore, the training dataset may be subject to intentional manipulations (e.g., data poisoning attacks \cite{papernot2018sok} and backdoor injections \cite{chen2017targeted}). Therefore, forensic investigations can benefit from access to the dataset and knowledge of its compilation methodology. Also, access to the modification history of the dataset can help with identifying intentional manipulations, as well as the responsible parties.
\subsubsection{Environment Forensics:}
The analogue of training data for reinforcement learning agents is the training environment. Similar to the case of training data, unrepresentative or manipulated environments may result in faulty behavior and failures (e.g., \cite{behzadan2018emergence} and \cite{behzadan2019sequential}). Hence, access to the training environment and its modification history can prove useful in forensic investigations 

\subsection{AI Substrate Forensics}
Substrate refers to the hardware and software platform that hosts the AI system. Forensic investigations of the AI substrate is essentially the domain of Digital Forensics. However, there are certain aspects of AI substrates that may give rise to unique circumstances. For instance, random bit-flips in the processor or memory due to cosmic rays has been shown to result in potentionally significant failures \cite{santoso2019understanding}. Furthermore, in cyber-physical AI systems, impaired or manipulated actuation of mechanical components (e.g., robotic locomotion) may result in misrepresentations of states and consequences of actions \cite{behzadan2018faults}. Main sources of forensic evidence in the AI substrate include the disk and memory, the network component, as well as the conditions of sensors and actuators in the cyber-physical settings. 

\subsection{AI Application Forensics}
AI is often deployed as a component of an application system. For instance, cloud-based cognitive services provide Application Programming Interfaces (APIs) to enable the integration of an AI service in software products. Forensic evidence collected from the usage logs of such APIs may indicate manipulation attempts, as well as failures in correct data cleaning and processing of queries. Furthermore, analysis of artefacts such as system resource utilization logs, authentication logs, and file system logs may also provide useful information on anomalous behavior and its causes. 

\subsection{AI Model Forensics}
Deployed machine learning models may result in failures that are independent of the aforementioned sources. For instance, original models may have been manipulated or replaced at some point in the machine learning supply chain. This type of malicious behavior may manifest in the form of backdoored models (e.g., \cite{chen2017targeted}), corrupted models (e.g., poisoning of malware classification \cite{chen2018automated}), or intentionally malicious models. While the domain of explainable AI offers an array of tools that may prove helpful in the analysis of the decision-making process in such models, a comprehensive forensic investigation requires further evidence which may be obtained from alternative sources, as discussed below: 
\subsubsection{Model Authentication Forensics:}
The aim of such evidence is to enable the verification of the authenticity of the model under investigation. Recent advances in watermarking techniques for machine learning models \cite{behzadan2019sequential,zhang2018protecting} provide the means for direct authentication of models. However, such approaches are yet to be commonly adopted. Furthermore, watermarks may also be prone to tampering and forging. Hence, alternative evidence such as software-level hashing techniques can provide a more reliable alternative.

\subsubsection{Model Identification / Extraction Forensics:}
If the model under investigation is a blackbox (i.e., model parameters and architecture are unknown), techniques such as model inversion and extraction \cite{tramer2016stealing} may provide the means for replicating its behavior for further testing and analysis. However, due to the approximate nature of such replicas, there remains the need for techniques that enable the quantification of uncertainty to maintain the legal soundness of the resulting forensic analyses. 

\subsubsection{Model Ballistics Forensics:}
In the general domain of Forensics, ballistics refers to the analysis and identification of the type and owner of a weapon used in a shooting incident. Similarly, in the forensic investigation of suspicious or malicious models, it is of importance to determine the type and creators (tools and individuals) of the model.

\subsubsection{Model Performance Forensics}
Recording the values of internal metrics and variables in the model may provide a detailed insight into the inner workings of the model. For instance, \cite{chen2018detecting} demonstrate that the analysis of activation values in deep learning models facilitates the detection of hidden backdoors. Also, higher-level measurements of model internals, such as state-action value estimates of reinforcement learning agents, and the multi-class probability distribution of classifiers, may provide useful evidence on the origins of failures. 

\subsubsection{Model Malware Forensics}
As mentioned before, machine learning models can be infected with backdoors and trigger-activated policies, or have been intentionally trained to act maliciously. An AI forensic investigation needs to detect and establish the existence of such malware, and provide the means to determine their malicious intent. Inspired the sandboxing techniques of computer malware analysis \cite{greamo2011sandboxing}, a preliminary source of forensic evidence in such cases is to replicate the conditions in simulation or a controlled environment to observe and analyze the behavioral dynamics of the model under investigation.


\section{Challenges}
\subsection{Unexplainability of AI}
Sound and indisputable root-cause analysis of failures in AI may require transparent and accurate interpretations of the decision-making process which resulted in undesired behavior. However, the research on the explainability of complex AI systems is still at its early stages, and the state of the art is far from solving the problem of explainability. Furthermore, some recent literature (e.g., \cite{yampolskiy2019unexplainability}) argue that as the AI technology and capabilities advance over time, it may become more difficult, or even impossible for AI systems to be explainable. In such circumstances, simpler abstractions of the decision-making process may enhance the forensic analysis of such failures. For instance, \cite{behzadan2018psychopathological} propose a psychopathological abstraction for complex AI safety problems. Similar abstractions may be required to enable accurate forensic analysis of advanced AI.
\subsection{AI Anti-Forensics}
In the domain of digital forensics, criminals constantly adapt to the state of the technology, and utilize techniques such as decoys, false evidence, or forensic cleaning to impede the forensic investigation.  It is likely that such anti-forensics techniques may also be invented and adopted by criminals to manipulate AI forensic investigations. Proactive identification of such techniques and development of mitigating solutions will thus become an increasingly important area of research in this domain. A recent anti-forensics general taxonomy was devised by researchers \cite{antifor}, yet, AI anti-forensics was not included in the taxonomy. While this is a challenge, it also presents a ripe opportunity for researchers. 
\subsection{Disconnect Between the Cyber Forensics and AI Communities}
One of the biggest challenges we face is the disconnect between the AI Safety and Cyber Forensics communities. Scientists from those two domains are not working together, thus, the domain of AI Forensics has not been conceived and is ripe for future work. This disconnect was apparent in a recent survey study, where the majority of digital forensic practitioners (67\%) (disagreed, agreed or were neutral) on their competency in Data Science \cite{AINeed2}.


\section{Conclusion}
We argued that the widespread integration of AI in everyday and critical technologies is bound to result in increased instances of failure, which will require technical investigations that produce legally acceptable and scientifically indisputable findings and conclusions. Inspired by the domain of cyber forensics, we thus introduced the need for the establishment of AI forensics as a new discipline under AI safety. Furthermore, we proposed a taxonomy of the subfields under this discipline, and presented a discussion on the foundational challenges that lay ahead of this new research area. 

\bibliography{AIForensics}
\bibliographystyle{aaai}

\end{document}